\documentclass{article}
\usepackage{spconf,graphicx}
\usepackage{amsmath,amssymb,amsfonts}
\usepackage{multirow}
\usepackage{booktabs}
\usepackage{xcolor,colortbl}
\usepackage{dblfloatfix}
\usepackage{caption}

\newcommand{\vect}[1]{\boldsymbol{#1}}

\newcommand{\HH}[1]{\ignorespaces}
\newcommand{\PG}[1]{\ignorespaces}
\newcommand{\KL}[1]{\ignorespaces}

\title{Continual learning using lattice-free MMI for speech recognition}
\ninept

\name{
      Hossein Hadian,
      Arseniy Gorin
}
\address{Behavox, Montreal, Canada}

\begin{document}

\maketitle

\begin{abstract}
Continual learning (CL), or domain expansion, recently became a popular topic  for automatic speech recognition (ASR) acoustic modeling because practical systems have to be  updated frequently in order to work robustly on  types of speech  not observed during initial training. While sequential adaptation allows tuning a system to a new domain, it may result in performance degradation on the old domains due to  catastrophic forgetting.
In this work we explore regularization-based CL for   neural network acoustic models trained with the  lattice-free  maximum mutual information (LF-MMI) criterion.
We simulate domain expansion by incrementally adapting the acoustic model on different public datasets that include several accents and speaking styles.
We investigate two well-known CL techniques, \emph{elastic weight consolidation} (EWC) and \emph{learning without forgetting} (LWF), which aim to reduce forgetting by preserving model weights or network outputs.
We additionally introduce a sequence-level LWF regularization, which exploits posteriors from the denominator graph of LF-MMI to further reduce forgetting.
Empirical results show that the proposed sequence-level LWF can improve the best average word error rate across all domains by up to  9.4\% relative compared with using regular LWF.

\end{abstract}

\section{Introduction}
\label{sec:intro}

Commercial automatic speech recognition (ASR) systems must frequently work robustly with a variety of accents, recording devices, background noises, and speech styles.
While training the acoustic models on thousands~\cite{likhomanenko2020rethinking,chan2021speechstew} and hundreds of thousands~\cite{narayanan2018toward} of hours of diverse speech helps the model generalize better, it still needs to be updated from time to time to cover domains that have not been sufficiently represented in the training data. Similar to~\cite{ghorbani2019domain} we define domain as a group of  utterances that share some common characteristics.
Retraining the model with new data from scratch is expensive and difficult if privacy limitations do not allow keeping target domain data and the initial training data in one place.
In practice, adaptation of the acoustic models using target domain speech appears to be a more favorable approach~\cite{bell2020adaptation}.
One problem is that the adapted model's performance on previously learned data may significantly degrade due to
catastrophic forgetting~\cite{mccloskey1989catastrophic,french1999catastrophic}, which limits the model's constant improvement across multiple domains.

Continual learning (CL) or life-long learning~\cite{parisi2019continual}, which was intensively studied especially in computer vision, is a variety of machine learning methods that aim to learn new tasks without forgetting previously acquired knowledge. Broadly, CL methods are based on regularization techniques~\cite{li2017learning,zenke2017continual,kirkpatrick2017overcoming}, dynamic neural network architectures~\cite{hayes2019memory,rusu2016progressive,yoon2017lifelong}, and rehearsal (also called memory replay)~\cite{hayes2019memory,lopez2017gradient,hinton1987using}.

Overcoming catastrophic forgetting when adapting the ASR acoustic models to new domains can be seen as a special case of CL, as the model usually keeps doing the same task (predicting the same set of classes).
Authors of~\cite{ghorbani2019domain} defined this task as domain expansion and explored several CL regularization techniques to optimize the ASR performance on initial and adapted speech accents. Later,~\cite{houston2020continual} explored similar regularization techniques to sequentially train a multi-dialect ASR system.
Alternatively,~\cite{sadhu2020continual} proposed a model expansion–based approach, which demonstrated good results but at a price of growing the model size with the number of training tasks, where tasks are defined as public datasets used by the authors. Similarly,~\cite{fainberg2021robust} explored a combination of model parameters and network outputs as a method for longitudinal training – a scenario in which the model learns a single domain but with incrementally appearing data and without access to historical data.
\cite{chang2021towards} compared regularization- and rehearsal-based techniques for incrementally learning an end-to-end CTC-based ASR system. They demonstrated better performance of rehearsal technique that, however, requires storing some portion of the training data from each domain.

Similar to~\cite{houston2020continual}, we focus on two popular regularization-based techniques, namely, elastic weight consolidation (EWC)~\cite{kirkpatrick2017overcoming} and learning without forgetting (LWF)~\cite{li2017learning}. We do not consider rehearsal-based techniques as in general they require access to the source data during adaptation and therefore are hard or impossible to use in privacy-preserving settings. Dynamic architectures are also not considered as these methods generally result in growing the model size over time, eventually slowing down the inference.

In this work\footnote{The source codes for this work will be released soon.} we apply EWC and LWF on acoustic models trained using the LF-MMI~\cite{povey2016purely}
criterion and assess their efficacy.
Furthermore, as the main contribution of this work, we propose a sequence-level version of LWF regularization that is based on LF-MMI and we demonstrate its effectiveness through multiple experiments.
The experiments are conducted on several public databases that involve several English accents (US, Indian, Singapore), speech styles (read speech, public talks, and conversations) and channels (studio and telephone).

\vspace{-3mm}
\section{Lattice-free MMI}
\label{sec:lfmmi}
Lattice-free MMI is a method of training an acoustic model using the MMI objective function
without generating
utterance-specific denominator lattices \cite{povey2016purely}.
More specifically, there is only one denominator graph that is used
for all utterances during training. The objective function is as follows\footnote{Note that in this and the rest of the equations in this paper,
for brevity we show the objective functions
for one utterance only. The full objective function would be the sum over all utterances. Also note that an objective function is \textbf{maximized} during training.}:
\begin{align}
  \label{eq:mmi_obj}
  \mathcal{F}_{MMI} &= \log \frac{  p_\lambda(\vect{x} | \mathbb{M}_{ \vect{w} }) }
                                                           {  p_\lambda(\vect{x})  }
\end{align}
The denominator can be approximated as:
\begin{align}
  \label{eq:mmi_den}
  p_\lambda(\vect{x}) &= \sum_{\vect{l}} p_\lambda(\vect{x} | \mathbb{M}_{ \vect{l} })
                        \approx p_\lambda(\vect{x} | \mathbb{M}_{den})
\end{align}
where $\mathbb{M}_{den}$ is a
finite-state transducer (FST) graph that includes all possible sequences of words.
This is called the denominator graph, as opposed to $\mathbb{M}_{ \vect{w} }$,
which is called the numerator graph and is utterance specific.
Expanding the log gives us:
\begin{align}
  \label{eq:mmi_obj2}
  \mathcal{F}_{MMI} &=  \log  p_\lambda(\vect{x} | \mathbb{M}_{ \vect{w} }) -
                                                           \log   p_\lambda(\vect{x} | \mathbb{M}_{den})
\end{align}
Note that the first term is maximizing the likelihood of the paths in the
numerator graph which represents
the true label and the second term is minimizing the likelihood
of all other paths in the denominator graph.

\vspace{-3mm}
\section{Continual learning}
\label{sec:cl}

The two most popular regularization-based approaches for CL (LWF and EWC) address the forgetting problem by forcing
the network to preserve either its outputs or its parameters, respectively. The following subsections briefly describe these approaches in more detail.

\subsection{Learning Without Forgetting}
\label{sec:lwf}
The LWF objective function is a cross-entropy term that is added as a regularization term to the main objective function for training the neural network:

\begin{equation}
\label{eq:lwf}
    \mathcal{F}_{LWF} = \alpha \sum_i  y^{src}_i \log(y_i)
\end{equation}
where $y_i$ are the outputs of the network while training and $y^{src}_i$ are the pre-computed
outputs of the network using the source model (i.e., the previous step of CL)
for all of the target domain training examples. We may simply call the $y^{src}_i$ reference posteriors in this context.
The scale $\alpha$ controls the effect of the LWF regularization.

\subsection{Elastic Weight Consolidation}
\label{sec:ewc}
While LWF concerns remembering the output distribution, EWC is focused on the parameters of the network. Specifically, the EWC objective function is as follows:
\begin{equation}
\label{eq:ewc}
    \mathcal{F}_{EWC} = - \alpha \sum_d \sum_j  F^d_j (\theta^d_j - \theta_j)^2
\end{equation}
where $d$ loops over previous domains, and $F^d_j$ is the Fisher diagonal
value for the $j_{th}$ parameter of
the network estimated using the data and model at CL step $d$.
Also $\theta_j$ is the $j_{th}$ parameter of
the current network and $\theta^d_j$ is the corresponding parameter from the model at CL step $d$.
Note that
similar to \cite{houston2020continual} we use
the online computation approach for the Fisher matrix estimation.
Also note that while LWF relies only on the output of CL at the previous domain, EWC needs
statistics (i.e., Fisher matrices) from all previous steps.

\vspace{-3mm}
\section{Proposed sequence level objective}
\label{sec:seq_lwf}
As mentioned earlier, the LWF loss is cross-entropy, which is frame-level. However, LF-MMI is a sequence-level loss function \cite{povey2016purely}.
It appears sub-optimal to train the acoustic model with a sequence-level objective and apply LWF using a frame-level objective. Therefore, instead
of using the neural network outputs as reference posteriors for LWF, we propose to use the denominator
posteriors (i.e., occupancies) as the reference posteriors. This leads to the following term for the LWF objective function:

\begin{equation}
\label{eq:bad_den}
    \mathcal{F} = \alpha \sum_i  \gamma^{src}_i \log(y_i)
\end{equation}
where $\gamma^{src}_i$ are the denominator posteriors from the previous domain. These posteriors are computed by running the forward-backward algorithm on the denominator graph given the neural network outputs and are therefore sequence-level. However, this cross-entropy term poses a problem because the
target distribution (i.e.,  $\gamma^{src}$) is sequence-level whereas the predicted distribution
(i.e., the neural network outputs while training: $y_i$) is frame-level. This results in excessive boosting
of the denominator posteriors as we are treating them as targets.\footnote{Note that if we try to train the network with this term, catastrophic forgetting is not prevented (results are not shown).} To offset this effect, we subtract the denominator log-prob term scaled by the same LWF scale $\alpha$.
This gives us the proposed sequence-level LWF objective function:
\begin{equation}
    \mathcal{F}_{DenLWF} = \alpha \sum_i \gamma^{src}_i \log(y_i) -
    \alpha \log  p_\lambda(\vect{x} | \mathbb{M}_{den})
\end{equation}
Taking the derivative of this term with respect to log outputs leads to the following term:
\begin{equation}
    \frac{\partial \mathcal{F}_{DenLWF}}{\partial \log(y_i)} = \alpha (\gamma^{src}_i - \gamma_i)
\end{equation}
where the $\gamma_i$ are the denominator occupancies during training. We can see immediately from
this equation that the gradients are $0$ at the beginning of
training since the training denominator
posteriors match the source ones. This is a desired characteristic, as in the beginning there
is no forgetting so we do not expect any gradients from the LWF regularization term.
The same cannot be said about Equation~\ref{eq:bad_den}, which leads to an unbalanced derivative.

\vspace{-2mm}
\section{Experimental setup}
\label{sec:exp_setup}

We choose 5 different English datasets (summarized in Table~\ref{tbl:data}) for sequential training.

\begin{table}[h]
   \centering
   \setlength\tabcolsep{2pt}
     \begin{tabular}{p{0.09\linewidth} | p{0.45\linewidth} | p{0.105\linewidth} | p{0.18\linewidth}| p{0.075\linewidth}}
      Name & Dataset & Speech type & English ~~ accent & Size, hours  \\
     \hline
     FSH & Fisher English~\cite{cieri2004fisher} subset & CTS\footnote{Conversational Telephony speech} & US &  200 \\
     \hline
     LIB & Librispeech~\cite{panayotov2015librispeech} subset & Read & Mostly US & 50  \\
      \hline
     TED & TED-LIUM~\cite{rousseau2014enhancing} subset & Talks  & Mostly US & 50 \\
      \hline
     NSC & National Speech Corpus~\cite{koh2019building} (IRV subset) & CTS & Singaporean & 50 \\
      \hline
     CVI & Mozilla Common Voice~\cite{commonvoice:2020} v6.1 (Indian speakers) & Read & Indian & 40 \\
      \hline
     \end{tabular}
   \caption{Summary of datasets used in the experiments.}
   \label{tbl:data}

 \end{table}

A 50-hour subset from each domain is used for domain expansion, except for CVI, for which there is only 40 hours of speech available. The 200-hour FSH dataset is used to train the seed model in all experiments. For each dataset, evaluation is done on their official dev/test subsets. For LIB, a combination of dev\_clean and test\_clean subsets is used and for TED, a combination of dev and test subsets.
A 3-gram language model is estimated using the training text of all domains and is used in
all experiments. We do not use i-vectors or other kinds of speaker adaptation in training the acoustic models.
In the following experiments, $FT$ refers to a model that has been trained using
naive fine-tuning and therefore is considered as the worst-case baseline
of catastrophic forgetting. On the other hand, $Comb$ refers to a model trained on all the available data from all domains combined \footnote{Note that unlike \cite{houston2020continual}, we do not train separate $comb$ models at each step as there is negligible difference in WER.}. Table~\ref{tbl:comb} summarizes
the WERs for the seed and $comb$ models on all domains. It is clear that
the most challenging domains for this seed model are CVI and NSC.

Kaldi ASR toolkit is used for all experiments~\cite{povey2011kaldi}.
The network is made up of 14 TDNN-F layers with a total of 14M parameters \cite{povey2018semi} and is trained using LF-MMI with mini-batch gradient descent.
The seed model's decision tree is used during all steps of domain expansion. The alignments at
each step are generated using a Gaussian mixture model (GMM) trained on the target domain at that step.

\begin{table}
\setlength{\belowcaptionskip}{-15pt}
   \centering
   \setlength\tabcolsep{3pt}
     \begin{tabular}{l c c c c c c}
      Model & FSH & TED & LIB & NSC & CVI & Avg \\
     \hline
Seed (FSH) & 18.2 & 15.6 & 11.8 & 56.3 & 58.4 & 32.1 \\
      \hline
Comb & 18.1 & 11.8 & 8.3 & 25.3 & 25.0 & 17.7 \\
      \hline
     \end{tabular}
   \caption{Results for the seed model and the combined-data model.}
   \label{tbl:comb}
 \end{table}

\definecolor{Gray}{gray}{0.85}
\newcolumntype{a}{>{\columncolor{Gray}}c}
\begin{table*}[hbt!]
	\renewcommand{\arraystretch}{1.3}
	%\vspace{0.3cm}
	\centering
	 \setlength\tabcolsep{2pt}
	\begin{tabular}{l   c c cca c ccca c cccca c ccccca}
	%	\topRule
		 & FSH & & \multicolumn{3}{c}{ $\rightarrow$ TED} & & \multicolumn{4}{c}{ $\rightarrow$ LIB}  & & \multicolumn{5}{c}{ $\rightarrow$ NSC} & & \multicolumn{6}{c}{ $\rightarrow$ CVI} \\
		\cline{2-2} \cline{4-6} \cline{8-11} \cline{13-17} \cline{19-24}
         Method & FSH & & FSH & TED & Avg & & FSH & TED & LIB & Avg & & FSH & TED & LIB & NSC & Avg & & FSH & TED & LIB & NSC & CVI & Avg \\
		\hline
	    FT         &  18.2 & & 30.6&13.3&21.9  & &  29.0&16.7&9.0& 18.2  & &  48.4&42.3&33.9&27.6&38.0  & &  59.5&42.2&23.5&58.0&27.9&42.2 \\
	    \hline
	    EWC         & 18.2 & & 21.7 & 13.5& 17.6  & &  22.1&14.4&9.7& 15.4  & &  29.9&20.6&15.5&34.0& 25.0  & &  38.4&25.3&15.4&47.0&31.7& 31.6 \\
		\hline
	    LWF         &  18.2 & & 23.0&13.5&18.2  & &  22.8&14.0&9.2&15.3  & &  30.5&18.3&12.5&30.5&23.0  & &  39.7&23.7&14.0&41.7&30.0&29.8 \\
	    \hline
	    \hline
	    DenLWF         &  18.2 & & 20.7&13.4&17.1  & &  22.5&13.9&9.1&15.2  & &  27.5&16.2&10.9&34.5&22.3  & &  33.0&19.7&12.0&40.0&30.3&27.0 \\
	    \hline
	\end{tabular}
	\caption{WERs,\% of 4 CL steps using different baseline and proposed methods. The first column shows the seed model. The next 3 columns under ``$\rightarrow TED$'' are for the first step FSH $\rightarrow$ TED, the columns under ``$\rightarrow LIB$'' are for the second step TED $\rightarrow$ LIB, and so forth.}
	\label{tbl:full}
\end{table*}

Similar to \cite{houston2020continual}, we also compute gap recovery rates for the CL steps as follows:
\begin{equation}
\label{eq:gap_recovery}
    \text{Gap Recovery} = 1.0 - \frac{\text{WER}_{CL} - \text{WER}_{Comb}}{\text{WER}_{FT} - \text{WER}_{Comb}}
\end{equation}
where $\text{WER}_{CL}$, $\text{WER}_{Comb}$, and $\text{WER}_{FT}$ refer
to WERs of the model trained using the CL method, the $comb$ model,
and the $FT$ model, respectively.
In addition to gap recovery rates, at each step we also define \emph{Relative Forgetting}, which  measures the degradation of WER on the past domains as a result of that step, and \emph{Relative Learning}, which measures the improvement of WER on the target domain at the end of that step.
Formally, they are defined as follows:
\begin{align}
\label{eq:rel_learning}
    \text{Relative Learning} &= 1.0 - \frac{\text{CWER}}{\text{CWER}^{src}} \\
\label{eq:rel_forgetting}
    \text{Relative Forgetting} &=  \frac{\text{PWER}}{\text{PWER}^{src}} - 1.0
\end{align}
where $\text{CWER}$ is the WER on the current
domain at each step (e.g., at step TED $\rightarrow$ LIB, it
would be the WER on the LIB test set at the end of the step).  $\text{CWER}^{src}$ is the WER on the same dataset, but using the model at the end of the previous step.
$\text{PWER}$ and $\text{PWER}^{src}$ are similar to $\text{CWER}$ and $\text{CWER}^{src}$ except they are measured on
past domains (averaged). For example,  at step TED $\rightarrow$ LIB of the sequence FSH $\rightarrow$ TED $\rightarrow$ LIB, the past domains are FSH and TED.

\vspace{-2mm}
\section{Experiments and Results}
\label{sec:exp}
This section describes a series of experiments to assess the efficacy of LWF, EWC, and
the proposed method on LF-MMI trained acoustic models. Specifically,
FSH is used to create the seed model, after which we do 4 steps of CL in the
following order: FSH $\rightarrow$ TED $\rightarrow$ LIB $\rightarrow$ NSC $\rightarrow$ CVI.
The detailed results are listed in Table~\ref{tbl:full}. Note that after each step, only the
domains involved in training up to that step are evaluated.
We can see that naive FT leads to the best performance on target data but severely degrades performance on all previous domains.
After the first step, EWC outperforms LWF. However, in the subsequent steps LWF performs considerably
better, especially in the last 2 steps. This is consistent with observations of \cite{houston2020continual}.
The results also show that DenLWF achieves
a WER similar to that of LWF in the second step and remarkably better WERs in
the other steps. Specifically, for the last two steps, which are the most challenging,
DenLWF relatively improves LWF's average WER by 3\% and 9.4\%, respectively. If we look at
the individual domain WERs, the improvements are even more visible. For example, in the last
step, while DenLWF achieves a slightly worse (but close) WER on CVI, it achieves a much better WER on previous domains. In particular, DenLWF improves LWF's WER on FSH from 39.7\% to 33.0\% and on TED from 23.7\% to 19.7\%.

It should be noted that all methods were tuned across multiple combinations of source-target transfers.
For LWF, we use a scale of $\alpha=1.0$ while for DenLWF a lower scale of $\alpha=0.6$ works best.
For EWC, after normalizing the Fisher diagonal values to
have a median of $1.0$, we use $\alpha=300$.

\begin{figure}[htb]
\setlength{\belowcaptionskip}{-5pt}

  \centering \includegraphics[width=0.45\textwidth]{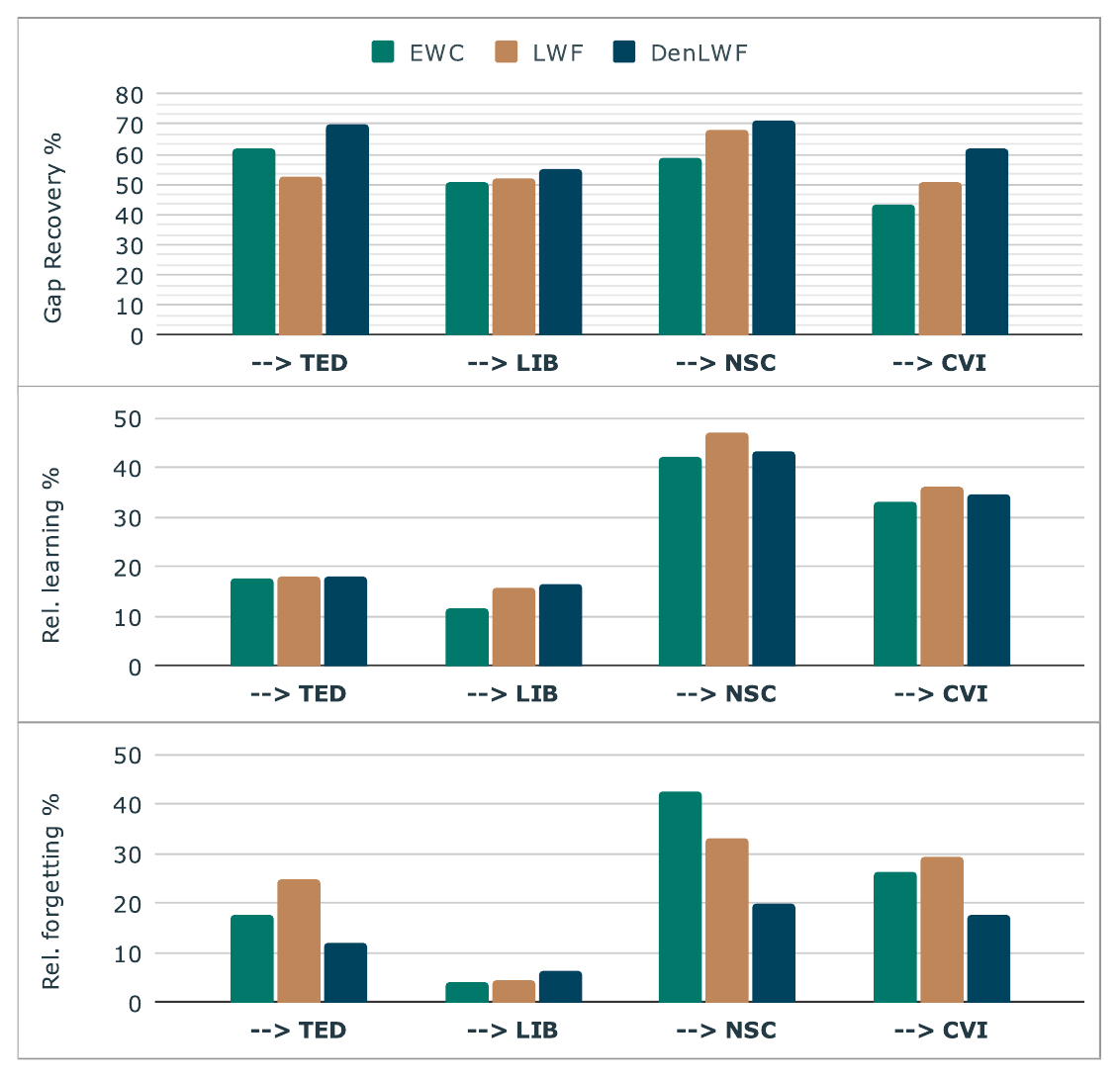}
  \caption{Overall performance of different methods. The upper chart shows average gap recovery percentages for different steps of the CL training pipeline. The middle and lower charts show relative learning and forgetting percentages, respectively.
  }
  \label{fig:overview}
\end{figure}

Figure~\ref{fig:overview} shows the gap recovery and relative learning/forgetting percentages (as defined in Equations \ref{eq:gap_recovery}-\ref{eq:rel_forgetting}) for the steps in Table~\ref{tbl:full}. In the top plot we can see that DenLWF achieves higher gap recovery rates in all steps, especially in the first and last steps. In the second step, all methods have a similar performance, which might be attributable to the fact that LIB is an easy domain for our seed model.
Specifically, EWC and LWF achieve a gap recovery rate of 40-60\% and 50-68\%, respectively, in different steps of CL, while DenLWF achieves a gap recovery rate of 55-71\% for these steps.
In the middle and lower charts we see the relative learning and forgetting rates, respectively. DenLWF and LWF show a similar capacity to learn new domains, with LWF learning slightly more in the third step. However, as can be seen in the lower chart, this comes at the cost of much more forgetting for LWF. Specifically, LWF has a relative forgetting rate of around 30\% in the last two steps whereas DenLWF only forgets around 20\% from the past domains.

\begin{figure}
\setlength{\belowcaptionskip}{-10pt}

  \centering \includegraphics[width=0.47\textwidth]{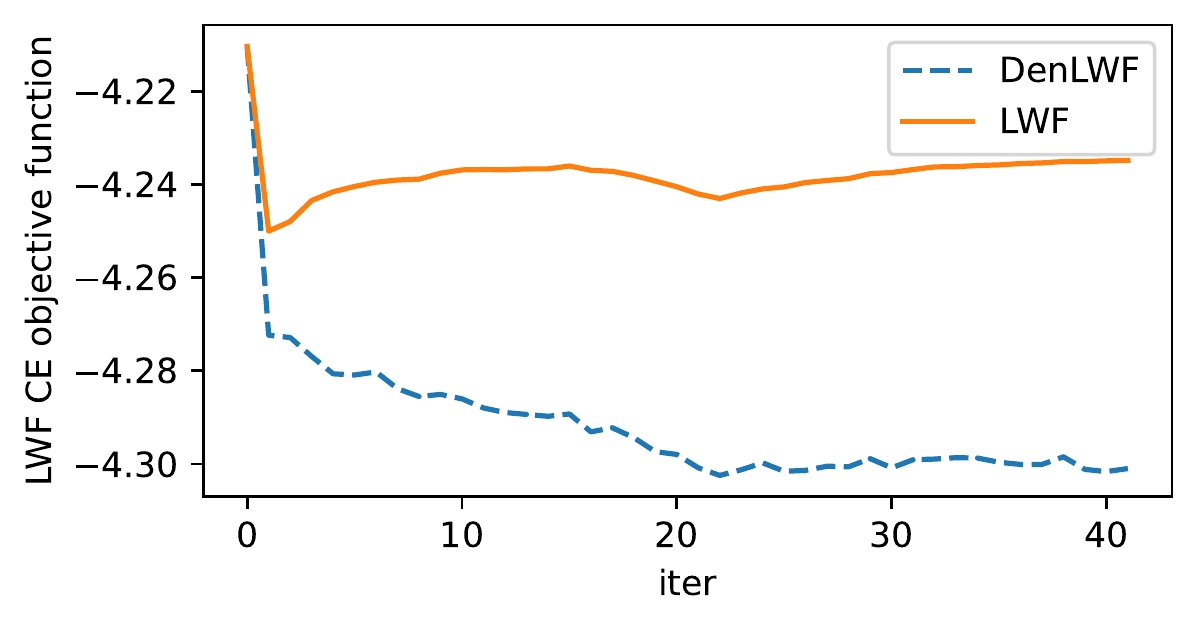}
  \caption{Comparing LWF cross-entropy objective during training for regular (frame-level) LWF and proposed DenLWF.}
  \label{fig:ce_loss}
\end{figure}

\subsection{Analysis and Discussion}
\label{sec:analysis}
In order to look more closely into how the proposed DenLWF functions and how it compares with regular LWF, we do two experiments. In the first experiment, we measure and plot the regular LWF objective
function (i.e., Equation~\ref{eq:lwf}) when training a model using regular LWF as well as DenLWF.
Note that LWF cross-entropy objective for DenLWF is computed here only for visualization purposes and is not used in back-propagation.
The resulting plot is shown in Figure~\ref{fig:ce_loss}. It can be observed that for regular LWF, the objective function degrades
right after the first iteration of training and then starts
to improve as training progresses. This
effectively restricts the output distribution from diverging too
far from the source domain posteriors.
However, in the case of DenLWF, we can see that
the output posteriors keep diverging from the reference
posteriors until halfway through training, at which point they plateau.
This seems to allow the network to
learn better, as the network output is under weaker frame-level restriction compared to regular LWF.
In other words, regular LWF forces the network to remember the source domain on a frame-level basis, whereas the proposed DenLWF objective forces the network to remember the source domain on a sequence-level basis -- which is more consistent with the speech signal -- leading to better overall performance (i.e., average WER).

\begin{figure}
\setlength{\belowcaptionskip}{-15pt}
  \centering \includegraphics[width=0.47\textwidth]{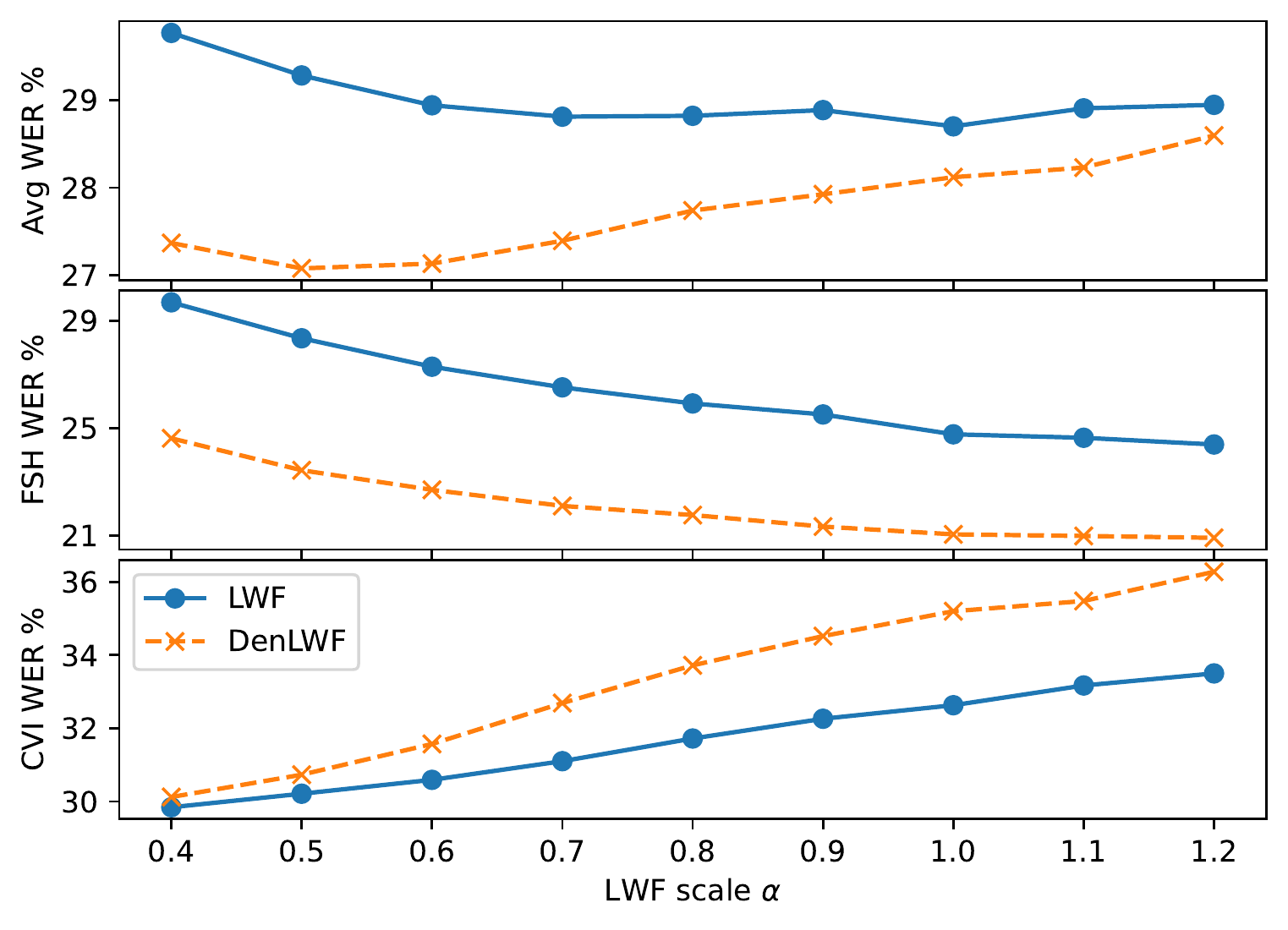}
  \caption{Average and individual WERs\% for regular LWF and DenLWF on FSH $\rightarrow$ CVI transfer.}
  \label{fig:avg_wer}
\end{figure}

In the second experiment, we investigate the impact of the LWF scale $\alpha$ on the average WER as well as on individual WERs. Figure~\ref{fig:avg_wer} compares regular LWF and DenLWF when trained using different scales’ $\alpha$ (all other hyperparameters are the same). The training step used in this experiment is FSH $\rightarrow$ CVI.
The upper plot shows the average WER (on FSH and CVI) while the lower plots show the WERs on FSH and CVI separately.
It can be seen that regular
LWF achieves its best performance when the scale is close to 1.0; whereas DenLWF achieves its best
performance when we use a smaller scale -- closer to 0.6. This suggests that DenLWF is a more
powerful regularizing term that needs a smaller scale.
We can also see that, generally, DenLWF achieves much less forgetting (i.e., lower WER on FSH), which comes at the cost of less learning (i.e., higher WER on CVI). However, at the scales where DenLWF works best
(i.e., $0.5$ and $0.6$), the difference in CVI WER between DenLWF and LWF is quite small.
Note that this plot also shows that for all the scales tested, the worst-case performance of DenLWF is better than the best-case performance of LWF on the FSH $\rightarrow$ CVI step. We observed similar plots for other steps (not shown due to lack of space).

\vspace{-1mm}
\section{Conclusions}
\label{sec:conclusion}
In this study, we applied two well-known regularization-based methods for continual learning, namely, EWC and LWF, to the state-of-the-art lattice-free MMI acoustic models and investigated their efficacy through a series of experiments. We observed gap recovery rates of 40-60\% using EWC and 50-68\% using LWF, for different steps of CL. Compared to naive fine-tuning, EWC improved the final average WER (after all the steps of CL) from 42.2\% to 31.6\%, while LWF improved it to 29.8\%.
In addition, given the fact that lattice-free MMI is a sequence-level objective function, we
proposed a new LWF objective function called DenLWF where we use MMI's denominator posteriors as reference posteriors in LWF while adding an offsetting term.
Through experiments we showed that DenLWF improved the gap recovery rates to
55-71\% for different steps of CL. Also, DenLWF improved the final average WER from 29.8\% down to 27.0\%.

\subsubsection*{Acknowledgements}
The authors would like to thank Dr. Daniel Povey for his valuable suggestions, and Stephen Gaudet for proofreading the paper.

\vfill\pagebreak

\bibliographystyle{IEEEbib}
%\bibliography{ms}

\end{document}